\documentclass[prl,twocolumn,superscriptaddress,showpacs,preprintnumbers,amsmath,amssymb]{revtex4}

\usepackage{graphicx}
\usepackage{dcolumn}
\usepackage{bm}

\begin{document}
\title{Density-Matrix Renormalization Group Study \\ 
of Trapped Imbalanced Fermi Condensates}
\author{Masaki Tezuka}
 \email{tezuka.m.aa@m.titech.ac.jp}
\affiliation{Department of Physics, Tokyo Institute of Technology,
2-12-1 O-okayama, Meguro-ku, Tokyo 152-8551, Japan}
\author{Masahito Ueda}
\affiliation{Department of Physics, Tokyo Institute of Technology,
2-12-1 O-okayama, Meguro-ku, Tokyo 152-8551, Japan}
\affiliation{ERATO Macroscopic Quantum Control Project, JST, Tokyo 113-8656, Japan}
\date{\today}

\begin{abstract}
The density-matrix renormalization group is employed to investigate a harmonically-trapped
imbalanced Fermi condensate based on a one-dimensional attractive Hubbard model.
The obtained density profile shows a flattened population difference of spin-up and spin-down
components at the center of the trap,
and exhibits phase separation between the condensate and unpaired majority atoms
for a certain range of the interaction and population imbalance $P$.
The two-particle density matrix reveals that the sign of the order parameter changes
periodically, demonstrating the realization of
the Fulde-Ferrell-Larkin-Ovchinnikov phase.
The minority spin atoms contribute to the quasi-condensate up to at least $P \simeq 0.8$.
Possible experimental situations to test our predictions are discussed.
\end{abstract}

\maketitle

Over the past few years, two major breakthroughs have been achieved in fermionic superfluids of
tenuous atomic vapor:
BEC-BCS crossover \cite{2004PhRvL..92d0403R,2005PhRvL..95b0404P}
and imbalanced superfluidity \cite{2006Sci...311..492Z,2006Sci...311..503P}.
These subjects have been studied not just in atomic physics but also in diverse subfields of physics \cite{2004RvMP...76..263C} such as
condensed matter physics
\cite{2000JPCM...12L.471M,
PhysRevB.63.140511,
2002PhRvB..66m4503T,
2004EL.....66..833M}
and nuclear physics \cite{2001PhRvD..63g4016A,2005PhLB..611..137G,2005PhLB..627...89C}.  
Two major issues in imbalanced superfluidity are 
whether superfluidity disappears
at a particular value (the Chandrasekhar-Clogston (CC) limit \cite{CClimit}) of population imbalance, 
and in what parameter regime the Fulde-Ferrell-Larkin-Ovchinnikov (FFLO) phase \cite{PhysRev.135.A550,larkin1964iss} emerges.
The observation of the CC limit is currently under controversy, while the FFLO phase remains elusive \cite{NoteFFLO}
despite extensive theory literature \cite{
2001EL.....55..150C, 
MachidaMizushimaIchioka, 
2005PhRvA..72b5601C,
2006PhRvL..96f0401S,
2006PhRvL..96k0403K,
2006PhRvL..97u0402G,
LiuHuDrummond,
2007PhRvA..75f3601Y, 
2007PhRvL..98g0402O, 
2007PhRvL..98q0402V}. 

In this Letter, we address these issues by applying the density-matrix renormalization group (DMRG) \cite{DMRG_White, 2005RvMP...77..259S}
to a system of harmonically trapped fermions. 
We consider a trapped one-dimensional (1D) system,
which can exhibit BEC due to the cut-off of infrared divergence.
We find that in the ground state of an imbalanced Fermi system the pairing order parameter
undergoes a periodic sign change ---the hallmark of the FFLO state
\cite{feiguin:220508,2007arXiv0710.1353B,2007arXiv0712.3364R}.
We also show that, as the imbalance becomes greater,
the minority-spin atoms continues to contribute to the quasi-condensed state,
implying non-existence of the CC limit.

We consider a system of spin-1/2 fermions undergoing short-ranged interaction and
confined in a 1D harmonic potential whose characteristic size is $l$.
We take $l$ as the unit of length and
discretize the system by introducing a lattice of $L$ sites
with the lattice constant given by $d=2l/L$.
The transfer amplitude between neighboring sites,
$t=\hbar^2/2md^2$ ($m$ is the mass of the atom),
reproduces the energy dispersion of the free space
in the limit of small filling factor ($L\rightarrow\infty$).
The contact interaction $g_\textrm{1D}\delta(z_\uparrow-z_\downarrow)$ is
approximated by introducing an on-site interaction with coupling constant 
$U=g_\textrm{1D}/d$ between atoms in different spin states.

We use DMRG to calculate the on-site pair correlation function and
the two-body reduced density matrix (2BDM)
for the $L$-site Hubbard model with a harmonic on-site potential.
DMRG allows an efficient, numerically exact
treatment of many-body problems in 1D
by iterative truncation of the Hilbert space.
We retain $200$ truncated states per DMRG block
with the maximum truncation error of $10^{-5}$.

The Hamiltonian of our system is given by
\begin{eqnarray}
\hat H &=& -t\sum_{i=0,\sigma}^{L-2}(\hat c_{i+1,\sigma}^\dag
\hat c_{i,\sigma}+\mathrm{h.c.})
+ U \sum_{i=0}^{L-1} \hat n_{i,\uparrow}\hat n_{i,\downarrow}\nonumber\\
&+& V\sum_{i=0}^{L-1} [i - (L-1)/2]^2 \hat n_{i},
\label{eqn:Hubbard}
\end{eqnarray}
where
$\hat c_{i,\sigma}$ annihilates an atom at site $i$ in spin state $\sigma(=\uparrow, \downarrow)$,
$\hat n_{i,\sigma}\equiv \hat c_{i,\sigma}^\dag \hat c_{i,\sigma}$,
$\hat n_i\equiv \hat n_{i,\uparrow}+\hat n_{i,\downarrow}$,
and $V\equiv4A/L^2$ is determined from the depth $A$ of the potential.
We note that $A/t\propto L^{-2}$ and $U/t\propto L^{-1}$.
We calculate the ground state of Hamiltonian (\ref{eqn:Hubbard}) within
the particle-number sector 
of $N_\uparrow$ and $N_\downarrow$ atoms in
spin states $\uparrow$ and $\downarrow$, respectively.
The imbalance parameter is defined as
$P\equiv (N_\uparrow-N_\downarrow)/N$,
where $N\equiv N_\uparrow+N_\downarrow$.
The Fermi momentum at the trap center is
calculated from the averaged density $n_\sigma\equiv
\overline{\langle \hat n_{i,\sigma} \rangle} L/2l$ as
$k_{\textrm{F}\sigma}=n_\sigma\pi$,
where the overbar denotes the average over
$0.1L$ -- $0.2L$ neighboring sites.

The $s$-wave scattering length of atoms $a_{1D}$ in a 1D trap with 
radial width $a_\perp\equiv\sqrt{\hbar/\mu \omega_\perp}$,
where $\mu$ is half the atom mass
and $\omega_\perp$ is the radial trapping frequency,
is modified from the free-space scattering length
$a_{3D}$ \cite{1998PhRvL..81..938O}:
in the low-energy limit of incident atoms,
$a_\textrm{1D} = 
-(a_\perp^2/2a_\textrm{3D})\times(1-Ca_\textrm{3D}/a_\perp)$ with
$C\simeq 1.46$.
Thus the 1D effective interaction is described by
$U(z) = g_\textrm{1D}\delta(z)$, where
$g_\textrm{1D} = -\hbar^2/\mu a_\textrm{1D}$.
We take $A/t=6400/L^2$, which, in the case of
${}^{40}\mbox{K}$ atoms with $a_\perp=86~\mbox{nm}$ and
$\omega_z=2\pi\times256~\mbox{Hz}$ \cite{2005PhRvL..94u0401M},
gives $l=6.28~\mu\mbox{m}$;
$U/t=-800/L$, for example, corresponds to $a_\textrm{1D}=62.8~\mbox{nm}$.

\begin{figure}
\includegraphics[width=8.6cm]{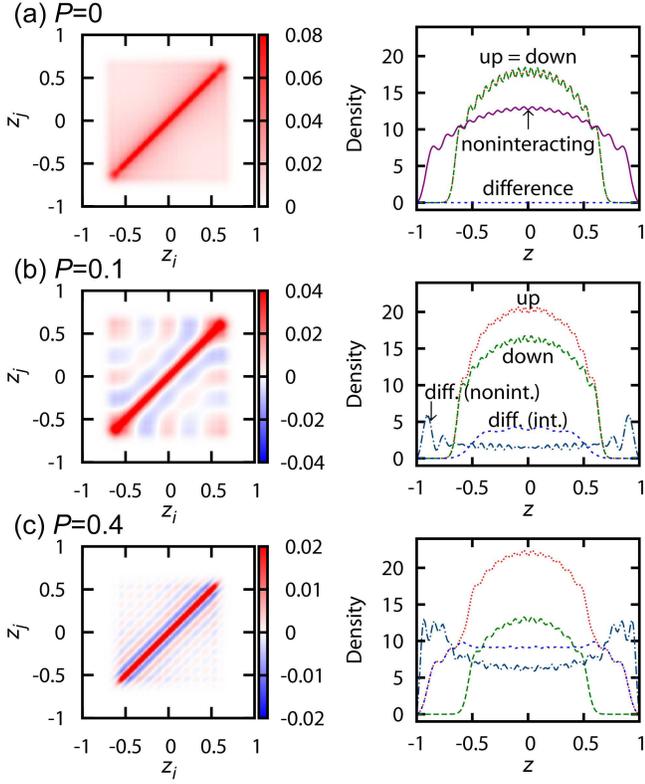}
\caption{
(Color online)
On-site pair correlation
$\langle \hat c_{i,\downarrow}^\dag \hat c_{i,\uparrow}^\dag
\hat c_{j,\uparrow} \hat c_{j,\downarrow}\rangle$ (left column)
and density distributions (right column)
of spin-up and spin-down atoms together with their difference
plotted against the $z$ coordinate in a harmonic potential
for $(A/t, U/t)=(6400/L^2,-800/L)$ and
$(N_\uparrow, N_\downarrow)=$
(a) $(20, 20)$, (b) $(22, 18)$ and (c) $(28, 12)$ with an $L=200$-site lattice.
In (a) the density of spin-up atoms, and in (b) and (c) the difference
for the noninteracting case ($U=0$) are also plotted.
}
\label{fig_FFLO3D}
\end{figure}

We first examine the on-site pair correlation function defined by
\begin{equation}
O_\mathrm{on-site}(z_i,z_j)\equiv
\langle \psi_0|\hat c_{i,\downarrow}^\dag \hat c_{i,\uparrow}^\dag
\hat c_{j,\uparrow} \hat c_{j,\downarrow}|\psi_0 \rangle,
\label{eqn:correlation}
\end{equation}
where $|\psi_0\rangle$ is the ground state of the system.
The left column of Fig.~\ref{fig_FFLO3D} displays $O_\mathrm{on-site}(z_i,z_j)$
for three values of $P$ with $L=200$ and
$U/t=-4$ [$(k_\textrm{F}a_\textrm{1D})^{-1}=1.82$ for $n_\uparrow=n_\downarrow=17.5$].
Figure~\ref{fig_FFLO3D}(a) exhibits the case with $P=0$,
where $O_\mathrm{on-site}(z_i,z_j)$ shows a slow decay
without changing the sign,
and then drops precipitously towards zero at $|z_i|\sim 0.75$.
The sharp boundary reflects cohesion of the system due to attractive interaction.
Figure~\ref{fig_FFLO3D}(b) shows the case with $P=0.1$, where
the pair correlation function changes its sign in real space,
and the amplitude of the oscillation vanishes rather sharply at
$|z_i|\sim 0.7$.
Figure~\ref{fig_FFLO3D} (c) shows the case with $P=0.4$,
where the sign change of the pair correlation function 
occurs with a much shorter period because of a larger separation in
the Fermi wave numbers of the up and down spin components.
The peaks of $O_\mathrm{on-site}(z_i,z_j)$ align along straight lines
of constant $z_i-z_j$, implying that the order parameter oscillates
periodically in space. 
In Figs.~\ref{fig_FFLO3D} (b) and (c),
the order parameter varies sinusoidally with 
no broken time-reversal symmetry,
indicating that the realized states are in
the Larkin-Ovchinnikov (LO) phase rather than the Fulde-Ferrell phase. 

\begin{figure}[t]
\includegraphics[width=8.6cm]{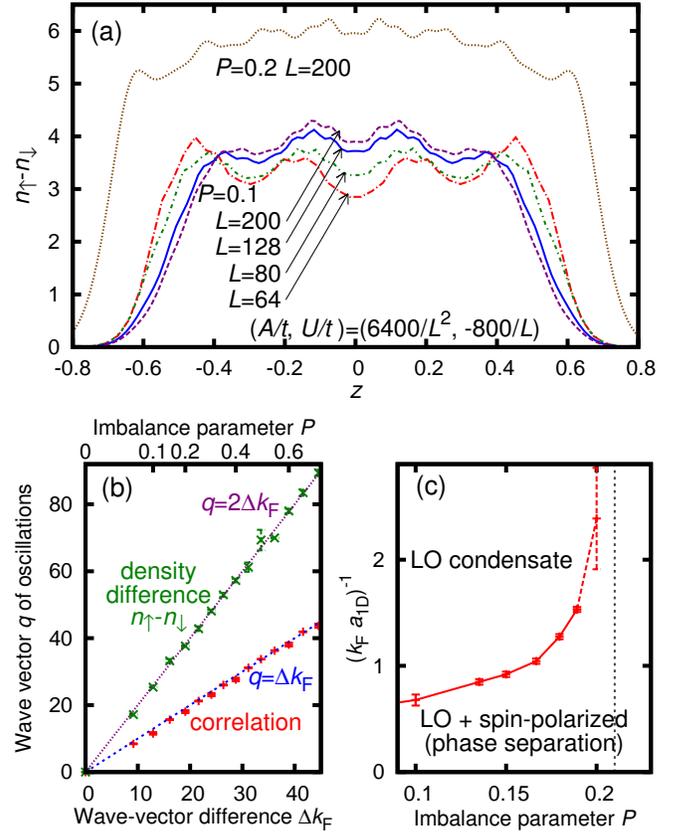}
\caption{
(Color online)
(a)
Dependence of the difference in density between spin-up
and spin-down atoms, $n_\uparrow-n_\downarrow$, on $z$
calculated for $N=40$ atoms with imbalance parameter $P=0.1$,
$L=64, 80, 128, 200$ and $(A/t,U/t)=(6400/L^2,-800/L)$.
$n_\uparrow-n_\downarrow$ for $P=0.2$ with $L=200$ is also plotted.
(b) Wave vector of oscillations in the pair correlation and
that in $n_\uparrow-n_\downarrow$ obtained for $N=40$ atoms
for various values of the imbalance parameter with $L=200$ and
$(A/t,U/t)=(0.16,-4)$.
(c) Phase diagram for $A/t=6400/L^2$.
\cite{LOphasediagram}
LO denotes the Larkin-Ovchinnikov state.
Here, $k_\textrm{F}$ is defined as
$(k_{\textrm{F}\uparrow}+k_{\textrm{F}\downarrow})/2$.
Phase separation occurs regardless of the value of $(k_\textrm{F}a_\textrm{1D})^{-1}$
when $P\geq 0.21$ (dotted line).
}
\label{fig_PHASE}
\end{figure}

The right column of Fig.~\ref{fig_FFLO3D} shows the density distributions
of the spin-up and spin-down atoms together with their difference.
For $P=0$, the total number of atoms peaks at the center of the harmonic trap with
small Friedel oscillation due to the presence of the trapping potential.
We note that the pair correlation extends over the whole region
where atoms are present.
In Fig.~\ref{fig_FFLO3D} (b) with $P=0.1$,
the difference in site occupation numbers is almost
flat near the trap center.
The population plummets again at the locations where the pair correlation function vanishes.
In Fig.~\ref{fig_FFLO3D} (c) with $P=0.4$, 
we see that the tails of the total population at both ends
only comprise the majority atoms in the $|\uparrow\rangle$ state.
The population difference first increases rapidly as we go from an edge
toward the center, and then becomes almost flat near the trap center.
The oscillating pair correlation in Fig.~\ref{fig_FFLO3D} (c) disappears at
$|z_i|\sim 0.6$ at which the number of the minority population vanishes.
This fact indicates that the two phases
---the FFLO condensate with almost constant density of excess majority atoms and the normal,
spin-polarized state--- phase-separate.

Figure~\ref{fig_PHASE} (a) plots the difference of density distributions of
spin-up and spin-down atoms for $P=0.1$ with varying $L$,
and for $P=0.2$ with $L=200$. We find that the density difference shows a
rapid convergence, which indicates that $L=200$ is close to the continuum limit,
and that the difference oscillates around a slowly-varying parabolic
curve around the center of the trap.
Near the trap center, the oscillation of the density difference
is incommensurate with the lattice and
almost independent of the lattice constant for $L\geq 80$.
By a nonlinear fit we determine
the wave vector of the oscillation as well as the density difference at
$z=0$ (the peak of the parabola).
Figure~\ref{fig_PHASE} (b) plots the wave vector of the oscillations against
the wave-vector difference $\Delta k_\textrm{F}\equiv
k_{\textrm{F}\uparrow}-k_{\textrm{F}\downarrow}$
for the pair correlation function and for the density difference
$n_\uparrow - n_\downarrow$ at the trap center.
The linear relation $q=\Delta k_\textrm{F}$, which holds for the former case,
is consistent with the LO phase \cite{feiguin:220508,2007arXiv0710.1353B,2007arXiv0712.3364R}.
The relation
$q=2\Delta k_\mathrm{F}$, which holds for the latter case,
is expected for the FFLO states \cite{1984PhRvB..30..122M},
and this is confirmed in Fig.~\ref{fig_PHASE} (b).

The sudden rise of the minority population for $P>0$ does
not appear in the noninteracting system and show striking resemblance to the
density profiles observed in the Rice experiment
\cite{2006PhRvL..97s0407P}.
Such cohesion in the density distribution implies
that the minority atoms are drawn into the inner core due to
pair correlation,
while unpaired atoms in the majority state are pushed outside the core. 
While similar density profiles have been reproduced in 3D simulations with
phenomenological surface tension
\cite{2006PhRvA..73e1602D,2007PhRvL..98z0406H},
it is interesting to observe
that such a steep rise also occurs in the 1D system by treating 
many-body effects rigorously.
We identify the point of phase separation with
the onset of the two shoulder peaks in the density difference.
Figure~\ref{fig_PHASE} (c) shows the phase diagram for $A/t=6400/L^2$
obtained by identifying the onset for $36\leq N \leq 41$.
As the population imbalance $P$ becomes greater,
stronger on-site attractive interaction or shorter $a_\textrm{1D}$
is needed to eliminate phase separation,
and for $P\geq0.21$, the shoulder peaks are always seen regardless
of the strength of the on-site attractive interaction.

\begin{figure}[t]
\includegraphics[width=8.6cm]{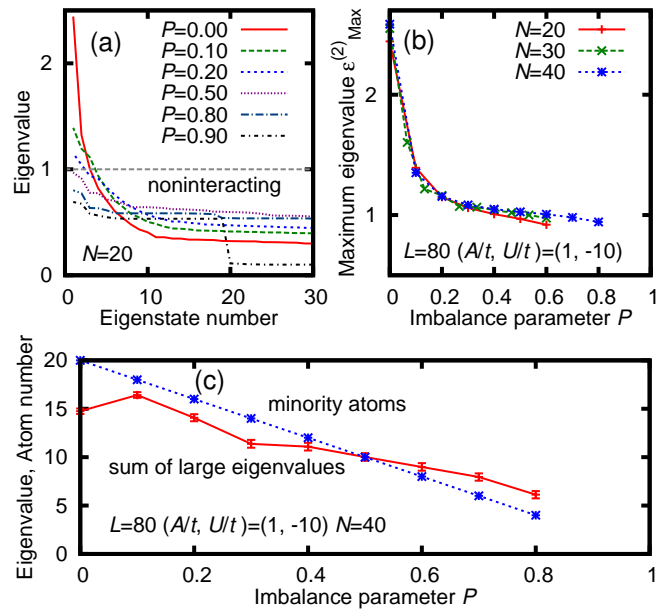}
\caption{
(Color online)
(a) Eigenvalue distribution for the two-particle density matrices
calculated for various values of the imbalance parameter $P$
for $(A/t, U/t)=(6400/L^2, -800/L)$ and $N=20$ on a $L=80$-site lattice.
(b) Maximum eigenvalue $\epsilon_\textrm{Max}^{(2)}$
for different total numbers of atoms $N=20,30,40$.
(c) The sum of eigenvalues up to the kink,
and the number of the minority spin atoms plotted against $P$.
}
\label{fig_2BDM}
\end{figure}

Finally, we study the dependence of the condensate fraction
on imbalance parameter $P$ by
calculating the eigenvalue distribution of the
two-body density matrix (2BDM) \cite{RevModPhys.34.694}.
The matrix elements of 2BDM in the site representation are given by
\begin{equation}
\rho_{ii',jj'}^{(2)}
\equiv
\langle\psi_0|\hat c_{i'\!\!,\downarrow}^\dag \hat c_{i,\uparrow}^\dag
\hat c_{j,\uparrow} \hat c_{j'\!\!,\downarrow}|\psi_0\rangle.
\end{equation}
Note that $\rho_{ii,jj}^{(2)}$ equals $O_{\rm on-site}(i,j)$.
The sum of the eigenvalues equals $N_\uparrow N_\downarrow$.
For the noninteracting case, there is an $N_\uparrow N_\downarrow$-fold
degenerate eigenvalue of unity, and the rest of the eigenvalues are all zero
for the ground state, because each single-particle energy level is either
occupied or unoccupied.

Figure~\ref{fig_2BDM} (a) shows the distribution of the eigenvalues
of 2BDM for various values of the imbalance parameter $P$.
We observe that the first few eigenvalues stand out from the rest.
This can be interpreted as the atoms forming a quasi-condensate,
where pairs of atoms condense into more than one eigenstate.
Figure~\ref{fig_2BDM} (b) shows how the maximum eigenvalue
decreases with increasing $P$.
It also shows that $\epsilon_\textrm{Max}^{(2)}$ is rather
insensitive to the value of $N$. This fact suggests that the lowest-lying
pairing state is already maximally occupied for a small value of $N$ and
that additional atoms contribute to other pairing states with oscillating
sign changes compatible with the LO state.
Figure~\ref{fig_2BDM} (c)
shows the sum of those large eigenvalues, together with the number
of the minority atoms.
We note that the sum is almost equal to the number of minority-spin atoms,
indicating that
almost all of the latter contributes to the quasi-condensate.
This phenomenon should be interpreted as being due to the abundance of majority-spin atoms
which can pair with minority-spin atoms no matter where the latter reside.
This behavior and the slow decay of the maximum eigenvalue of 2BDM
suggests the robustness of condensate against population imbalance.

The eigenvector $\phi_{ij}$ of the 2BDM with the largest eigenvalue
$\epsilon_{\rm Max}^{(2)}$, which describes the state in which
the largest number of Cooper pairs participate,
is contributed mostly from the $i=j$ components.
This reflects the fact that the two atoms on the same site are most 
effectively paired,
as expected from Hamiltonian (\ref{eqn:Hubbard}).
The sign of $\phi_{ii}$ changes as a function of $i$ when $P>0$,
showing that the sign change in the order parameter originates
from the FFLO nature of the condensate rather than from the Andreev
scattering of particles at the normal-superfluid boundary \cite{2007PhRvL..98q0402V}.

In conclusion, we have employed the density-matrix renormalization group
to investigate fermionic condensation of a 1D trapped gas with
population imbalance between the two spin states,
and have shown that the ground state exhibits the FFLO-like feature over
a wide range of imbalance parameter $P$.
We have also found a critical value of $P$ beyond which the LO condensate
and spin-polarized Fermi gas phase-separate at $T=0$ no matter
how strong the interaction is.

This work was supported by a Grant-in-Aid for Scientific
Research (Grant No. 17071005) and by a 21st Century COE program on ``Nanometer-Scale Quantum Physics''
from the Ministry of Education, Culture, Sports, Science
and Technology of Japan.
M.T. is supported by Research Fellowship
of the Japan Society for the Promotion of Science for Young Scientists.

\bibliography{imbalance}

\begin{thebibliography}{39}
\expandafter\ifx\csname natexlab\endcsname\relax\def\natexlab#1{#1}\fi
\expandafter\ifx\csname bibnamefont\endcsname\relax
  \def\bibnamefont#1{#1}\fi
\expandafter\ifx\csname bibfnamefont\endcsname\relax
  \def\bibfnamefont#1{#1}\fi
\expandafter\ifx\csname citenamefont\endcsname\relax
  \def\citenamefont#1{#1}\fi
\expandafter\ifx\csname url\endcsname\relax
  \def\url#1{\texttt{#1}}\fi
\expandafter\ifx\csname urlprefix\endcsname\relax\def\urlprefix{URL }\fi
\providecommand{\bibinfo}[2]{#2}
\providecommand{\eprint}[2][]{\url{#2}}

\bibitem[{\citenamefont{{Regal} et~al.}(2004)\citenamefont{{Regal}, {Greiner},
  and {Jin}}}]{2004PhRvL..92d0403R}
\bibinfo{author}{\bibfnamefont{C.~A.} \bibnamefont{{Regal}}},
  \bibinfo{author}{\bibfnamefont{M.}~\bibnamefont{{Greiner}}},
  \bibnamefont{and} \bibinfo{author}{\bibfnamefont{D.~S.} \bibnamefont{{Jin}}},
  \bibinfo{journal}{\prl} \textbf{\bibinfo{volume}{92}},
  \bibinfo{pages}{040403} (\bibinfo{year}{2004}).

\bibitem[{\citenamefont{{Partridge} et~al.}(2005)\citenamefont{{Partridge},
  {Strecker}, {Kamar}, {Jack}, and {Hulet}}}]{2005PhRvL..95b0404P}
\bibinfo{author}{\bibfnamefont{G.~B.} \bibnamefont{{Partridge}}},
  \bibinfo{author}{\bibfnamefont{K.~E.} \bibnamefont{{Strecker}}},
  \bibinfo{author}{\bibfnamefont{R.~I.} \bibnamefont{{Kamar}}},
  \bibinfo{author}{\bibfnamefont{M.~W.} \bibnamefont{{Jack}}},
  \bibnamefont{and} \bibinfo{author}{\bibfnamefont{R.~G.}
  \bibnamefont{{Hulet}}}, \bibinfo{journal}{\prl}
  \textbf{\bibinfo{volume}{95}}, \bibinfo{pages}{020404}
  (\bibinfo{year}{2005}).

\bibitem[{\citenamefont{{Zwierlein} et~al.}(2006)\citenamefont{{Zwierlein},
  {Schirotzek}, {Schunck}, and {Ketterle}}}]{2006Sci...311..492Z}
\bibinfo{author}{\bibfnamefont{M.~W.} \bibnamefont{{Zwierlein}}},
  \bibinfo{author}{\bibfnamefont{A.}~\bibnamefont{{Schirotzek}}},
  \bibinfo{author}{\bibfnamefont{C.~H.} \bibnamefont{{Schunck}}},
  \bibnamefont{and}
  \bibinfo{author}{\bibfnamefont{W.}~\bibnamefont{{Ketterle}}},
  \bibinfo{journal}{Science} \textbf{\bibinfo{volume}{311}},
  \bibinfo{pages}{492} (\bibinfo{year}{2006}).

\bibitem[{\citenamefont{{Partridge}
  et~al.}(2006{\natexlab{a}})\citenamefont{{Partridge}, {Li}, {Kamar}, {Liao},
  and {Hulet}}}]{2006Sci...311..503P}
\bibinfo{author}{\bibfnamefont{G.~B.} \bibnamefont{{Partridge}}},
  \bibinfo{author}{\bibfnamefont{W.}~\bibnamefont{{Li}}},
  \bibinfo{author}{\bibfnamefont{R.~I.} \bibnamefont{{Kamar}}},
  \bibinfo{author}{\bibfnamefont{Y.-a.} \bibnamefont{{Liao}}},
  \bibnamefont{and} \bibinfo{author}{\bibfnamefont{R.~G.}
  \bibnamefont{{Hulet}}}, \bibinfo{journal}{Science}
  \textbf{\bibinfo{volume}{311}}, \bibinfo{pages}{503}
  (\bibinfo{year}{2006}{\natexlab{a}}).

\bibitem[{\citenamefont{{Casalbuoni} and
  {Nardulli}}(2004)}]{2004RvMP...76..263C}
\bibinfo{author}{\bibfnamefont{R.}~\bibnamefont{{Casalbuoni}}}
  \bibnamefont{and}
  \bibinfo{author}{\bibfnamefont{G.}~\bibnamefont{{Nardulli}}},
  \bibinfo{journal}{\rmp} \textbf{\bibinfo{volume}{76}}, \bibinfo{pages}{263}
  (\bibinfo{year}{2004}).

\bibitem[{\citenamefont{{Manalo} and {Klein}}(2000)}]{2000JPCM...12L.471M}
\bibinfo{author}{\bibfnamefont{S.}~\bibnamefont{{Manalo}}} \bibnamefont{and}
  \bibinfo{author}{\bibfnamefont{U.}~\bibnamefont{{Klein}}},
  \bibinfo{journal}{J. Phys.: Cond. Mat.} \textbf{\bibinfo{volume}{12}},
  \bibinfo{pages}{L471} (\bibinfo{year}{2000}).

\bibitem[{\citenamefont{{Tanatar} et~al.}(2002)\citenamefont{{Tanatar},
  {Ishiguro}, {Tanaka}, and {Kobayashi}}}]{2002PhRvB..66m4503T}
\bibinfo{author}{\bibfnamefont{M.~A.} \bibnamefont{{Tanatar}}},
  \bibinfo{author}{\bibfnamefont{T.}~\bibnamefont{{Ishiguro}}},
  \bibinfo{author}{\bibfnamefont{H.}~\bibnamefont{{Tanaka}}}, \bibnamefont{and}
  \bibinfo{author}{\bibfnamefont{H.}~\bibnamefont{{Kobayashi}}},
  \bibinfo{journal}{\prb} \textbf{\bibinfo{volume}{66}},
  \bibinfo{pages}{134503} (\bibinfo{year}{2002}).

\bibitem[{\citenamefont{{Mora} and {Combescot}}(2004)}]{2004EL.....66..833M}
\bibinfo{author}{\bibfnamefont{C.}~\bibnamefont{{Mora}}} \bibnamefont{and}
  \bibinfo{author}{\bibfnamefont{R.}~\bibnamefont{{Combescot}}},
  \bibinfo{journal}{Europhys. Lett.} \textbf{\bibinfo{volume}{66}},
  \bibinfo{pages}{833} (\bibinfo{year}{2004}).

\bibitem[{\citenamefont{Yang}(2001)}]{PhysRevB.63.140511}
\bibinfo{author}{\bibfnamefont{K.}~\bibnamefont{Yang}}, \bibinfo{journal}{Phys.
  Rev. B} \textbf{\bibinfo{volume}{63}}, \bibinfo{pages}{140511}
  (\bibinfo{year}{2001}).

\bibitem[{\citenamefont{{Alford} et~al.}(2001)\citenamefont{{Alford}, {Bowers},
  and {Rajagopal}}}]{2001PhRvD..63g4016A}
\bibinfo{author}{\bibfnamefont{M.}~\bibnamefont{{Alford}}},
  \bibinfo{author}{\bibfnamefont{J.~A.} \bibnamefont{{Bowers}}},
  \bibnamefont{and}
  \bibinfo{author}{\bibfnamefont{K.}~\bibnamefont{{Rajagopal}}},
  \bibinfo{journal}{\prd} \textbf{\bibinfo{volume}{63}},
  \bibinfo{pages}{074016} (\bibinfo{year}{2001}).

\bibitem[{\citenamefont{{Giannakis} and {Ren}}(2005)}]{2005PhLB..611..137G}
\bibinfo{author}{\bibfnamefont{I.}~\bibnamefont{{Giannakis}}} \bibnamefont{and}
  \bibinfo{author}{\bibfnamefont{H.-C.} \bibnamefont{{Ren}}},
  \bibinfo{journal}{Phys. Lett. B} \textbf{\bibinfo{volume}{611}},
  \bibinfo{pages}{137} (\bibinfo{year}{2005}).

\bibitem[{\citenamefont{{Casalbuoni} et~al.}(2005)\citenamefont{{Casalbuoni},
  {Gatto}, {Ippolito}, {Nardulli}, and {Ruggieri}}}]{2005PhLB..627...89C}
\bibinfo{author}{\bibfnamefont{R.}~\bibnamefont{{Casalbuoni}}},
  \bibinfo{author}{\bibfnamefont{R.}~\bibnamefont{{Gatto}}},
  \bibinfo{author}{\bibfnamefont{N.}~\bibnamefont{{Ippolito}}},
  \bibinfo{author}{\bibfnamefont{G.}~\bibnamefont{{Nardulli}}},
  \bibnamefont{and}
  \bibinfo{author}{\bibfnamefont{M.}~\bibnamefont{{Ruggieri}}},
  \bibinfo{journal}{Phys. Lett. B} \textbf{\bibinfo{volume}{627}},
  \bibinfo{pages}{89} (\bibinfo{year}{2005}).

\bibitem[{CCl()}]{CClimit}
\bibinfo{note}{\bibinfo{author}{\bibfnamefont{B.~S.}
  \bibnamefont{Chandrasekhar}}, \bibinfo{journal}{App. Phys. Lett.}
  \textbf{\bibinfo{volume}{1}}, \bibinfo{pages}{7} (\bibinfo{year}{1962});
  \bibinfo{author}{\bibfnamefont{A.~M.} \bibnamefont{Clogston}},
  \bibinfo{journal}{\prl} \textbf{\bibinfo{volume}{9}}, \bibinfo{pages}{266}
  (\bibinfo{year}{1962}).}

\bibitem[{\citenamefont{Fulde and Ferrell}(1964)}]{PhysRev.135.A550}
\bibinfo{author}{\bibfnamefont{P.}~\bibnamefont{Fulde}} \bibnamefont{and}
  \bibinfo{author}{\bibfnamefont{R.~A.} \bibnamefont{Ferrell}},
  \bibinfo{journal}{Phys. Rev.} \textbf{\bibinfo{volume}{135}},
  \bibinfo{pages}{A550} (\bibinfo{year}{1964}).

\bibitem[{\citenamefont{Larkin and Ovchinnikov}(1964)}]{larkin1964iss}
\bibinfo{author}{\bibfnamefont{A.}~\bibnamefont{Larkin}} \bibnamefont{and}
  \bibinfo{author}{\bibfnamefont{Y.}~\bibnamefont{Ovchinnikov}},
  \bibinfo{journal}{Zh. Eksp. Teor. Fiz} \textbf{\bibinfo{volume}{47}},
  \bibinfo{pages}{1136} (\bibinfo{year}{1964}).

\bibitem[{Not()}]{NoteFFLO}
\bibinfo{note}{In condense matter evidence of the FFLO phase in superconductors
  has recently been presented:
  \bibinfo{author}{\bibfnamefont{A.}~\bibnamefont{{Bianchi}}},
  \bibinfo{author}{\bibfnamefont{R.}~\bibnamefont{{Movshovich}}},
  \bibinfo{author}{\bibfnamefont{C.}~\bibnamefont{{Capan}}},
  \bibinfo{author}{\bibfnamefont{P.~G.} \bibnamefont{{Pagliuso}}},
  \bibnamefont{and} \bibinfo{author}{\bibfnamefont{J.~L.}
  \bibnamefont{{Sarrao}}}, \bibinfo{journal}{\prl}
  \textbf{\bibinfo{volume}{91}}, \bibinfo{pages}{187004}
  (\bibinfo{year}{2003});
  \bibinfo{author}{\bibfnamefont{K.}~\bibnamefont{{Kakuyanagi}}},
  \bibinfo{author}{\bibfnamefont{M.}~\bibnamefont{{Saitoh}}},
  \bibinfo{author}{\bibfnamefont{K.}~\bibnamefont{{Kumagai}}},
  \bibinfo{author}{\bibfnamefont{S.}~\bibnamefont{{Takashima}}},
  \bibinfo{author}{\bibfnamefont{M.}~\bibnamefont{{Nohara}}},
  \bibinfo{author}{\bibfnamefont{H.}~\bibnamefont{{Takagi}}}, \bibnamefont{and}
  \bibinfo{author}{\bibfnamefont{Y.}~\bibnamefont{{Matsuda}}},
  \bibinfo{journal}{\prl} \textbf{\bibinfo{volume}{94}},
  \bibinfo{pages}{047602} (\bibinfo{year}{2005});
  \bibinfo{author}{\bibfnamefont{V.~F.} \bibnamefont{{Correa}}},
  \bibinfo{author}{\bibfnamefont{T.~P.} \bibnamefont{{Murphy}}},
  \bibinfo{author}{\bibfnamefont{C.}~\bibnamefont{{Martin}}},
  \bibinfo{author}{\bibfnamefont{K.~M.} \bibnamefont{{Purcell}}},
  \bibinfo{author}{\bibfnamefont{E.~C.} \bibnamefont{{Palm}}},
  \bibinfo{author}{\bibfnamefont{G.~M.} \bibnamefont{{Schmiedeshoff}}},
  \bibinfo{author}{\bibfnamefont{J.~C.} \bibnamefont{{Cooley}}},
  \bibnamefont{and} \bibinfo{author}{\bibfnamefont{S.~W.}
  \bibnamefont{{Tozer}}}, \bibinfo{journal}{\prl}
  \textbf{\bibinfo{volume}{98}}, \bibinfo{pages}{087001}
  (\bibinfo{year}{2007}).}

\bibitem[{\citenamefont{{Combescot}}(2001)}]{2001EL.....55..150C}
\bibinfo{author}{\bibfnamefont{R.}~\bibnamefont{{Combescot}}},
  \bibinfo{journal}{Europhys. Lett.} \textbf{\bibinfo{volume}{55}},
  \bibinfo{pages}{150} (\bibinfo{year}{2001}).

\bibitem[{Mac()}]{MachidaMizushimaIchioka}
\bibinfo{note}{\bibinfo{author}{\bibfnamefont{T.}~\bibnamefont{Mizushima}},
  \bibinfo{author}{\bibfnamefont{K.}~\bibnamefont{Machida}}, \bibnamefont{and}
  \bibinfo{author}{\bibfnamefont{M.}~\bibnamefont{Ichioka}},
  \bibinfo{journal}{\prl} \textbf{\bibinfo{volume}{94}}, \bibinfo{eid}{060404}
  (\bibinfo{year}{2005});
  \bibinfo{author}{\bibfnamefont{K.}~\bibnamefont{{Machida}}},
  \bibinfo{author}{\bibfnamefont{T.}~\bibnamefont{{Mizushima}}},
  \bibnamefont{and}
  \bibinfo{author}{\bibfnamefont{M.}~\bibnamefont{{Ichioka}}},
  \bibinfo{journal}{\prl} \textbf{\bibinfo{volume}{97}},
  \bibinfo{pages}{120407} (\bibinfo{year}{2006}).}

\bibitem[{\citenamefont{{Castorina} et~al.}(2005)\citenamefont{{Castorina},
  {Grasso}, {Oertel}, {Urban}, and {Zappal{\`a}}}}]{2005PhRvA..72b5601C}
\bibinfo{author}{\bibfnamefont{P.}~\bibnamefont{{Castorina}}},
  \bibinfo{author}{\bibfnamefont{M.}~\bibnamefont{{Grasso}}},
  \bibinfo{author}{\bibfnamefont{M.}~\bibnamefont{{Oertel}}},
  \bibinfo{author}{\bibfnamefont{M.}~\bibnamefont{{Urban}}}, \bibnamefont{and}
  \bibinfo{author}{\bibfnamefont{D.}~\bibnamefont{{Zappal{\`a}}}},
  \bibinfo{journal}{\pra} \textbf{\bibinfo{volume}{72}},
  \bibinfo{pages}{025601} (\bibinfo{year}{2005}).

\bibitem[{\citenamefont{{Sheehy} and
  {Radzihovsky}}(2006)}]{2006PhRvL..96f0401S}
\bibinfo{author}{\bibfnamefont{D.~E.} \bibnamefont{{Sheehy}}} \bibnamefont{and}
  \bibinfo{author}{\bibfnamefont{L.}~\bibnamefont{{Radzihovsky}}},
  \bibinfo{journal}{\prl} \textbf{\bibinfo{volume}{96}},
  \bibinfo{pages}{060401} (\bibinfo{year}{2006}).

\bibitem[{\citenamefont{{Kinnunen} et~al.}(2006)\citenamefont{{Kinnunen},
  {Jensen}, and {T{\"o}rm{\"a}}}}]{2006PhRvL..96k0403K}
\bibinfo{author}{\bibfnamefont{J.}~\bibnamefont{{Kinnunen}}},
  \bibinfo{author}{\bibfnamefont{L.~M.} \bibnamefont{{Jensen}}},
  \bibnamefont{and}
  \bibinfo{author}{\bibfnamefont{P.}~\bibnamefont{{T{\"o}rm{\"a}}}},
  \bibinfo{journal}{\prl} \textbf{\bibinfo{volume}{96}},
  \bibinfo{pages}{110403} (\bibinfo{year}{2006}).

\bibitem[{\citenamefont{{Gubbels} et~al.}(2006)\citenamefont{{Gubbels},
  {Romans}, and {Stoof}}}]{2006PhRvL..97u0402G}
\bibinfo{author}{\bibfnamefont{K.~B.} \bibnamefont{{Gubbels}}},
  \bibinfo{author}{\bibfnamefont{M.~W.~J.} \bibnamefont{{Romans}}},
  \bibnamefont{and} \bibinfo{author}{\bibfnamefont{H.~T.~C.}
  \bibnamefont{{Stoof}}}, \bibinfo{journal}{\prl}
  \textbf{\bibinfo{volume}{97}}, \bibinfo{pages}{210402}
  (\bibinfo{year}{2006}).

\bibitem[{Liu()}]{LiuHuDrummond}
\bibinfo{note}{\bibinfo{author}{\bibfnamefont{X.-J.} \bibnamefont{{Liu}}},
  \bibinfo{author}{\bibfnamefont{H.}~\bibnamefont{{Hu}}}, \bibnamefont{and}
  \bibinfo{author}{\bibfnamefont{P.~D.} \bibnamefont{{Drummond}}},
  \bibinfo{journal}{\pra} \textbf{\bibinfo{volume}{75}},
  \bibinfo{pages}{023614} (\bibinfo{year}{2007});
  \bibinfo{author}{\bibfnamefont{H.}~\bibnamefont{{Hu}}},
  \bibinfo{author}{\bibfnamefont{X.-J.} \bibnamefont{{Liu}}}, \bibnamefont{and}
  \bibinfo{author}{\bibfnamefont{P.~D.} \bibnamefont{{Drummond}}},
  \bibinfo{journal}{\prl} \textbf{\bibinfo{volume}{98}},
  \bibinfo{pages}{070403} (\bibinfo{year}{2007}).}

\bibitem[{\citenamefont{{Yoshida} and {Yip}}(2007)}]{2007PhRvA..75f3601Y}
\bibinfo{author}{\bibfnamefont{N.}~\bibnamefont{{Yoshida}}} \bibnamefont{and}
  \bibinfo{author}{\bibfnamefont{S.-K.} \bibnamefont{{Yip}}},
  \bibinfo{journal}{\pra} \textbf{\bibinfo{volume}{75}},
  \bibinfo{pages}{063601} (\bibinfo{year}{2007}).

\bibitem[{\citenamefont{{Orso}}(2007)}]{2007PhRvL..98g0402O}
\bibinfo{author}{\bibfnamefont{G.}~\bibnamefont{{Orso}}},
  \bibinfo{journal}{\prl} \textbf{\bibinfo{volume}{98}},
  \bibinfo{pages}{070402} (\bibinfo{year}{2007}).

\bibitem[{\citenamefont{{van Schaeybroeck} and
  {Lazarides}}(2007)}]{2007PhRvL..98q0402V}
\bibinfo{author}{\bibfnamefont{B.}~\bibnamefont{{van Schaeybroeck}}}
  \bibnamefont{and}
  \bibinfo{author}{\bibfnamefont{A.}~\bibnamefont{{Lazarides}}},
  \bibinfo{journal}{\prl} \textbf{\bibinfo{volume}{98}},
  \bibinfo{pages}{170402} (\bibinfo{year}{2007}).

\bibitem[{DMR()}]{DMRG_White}
\bibinfo{note}{\bibinfo{author}{\bibfnamefont{S.~R.} \bibnamefont{{White}}},
  \bibinfo{journal}{\prl} \textbf{\bibinfo{volume}{69}}, \bibinfo{pages}{2863}
  (\bibinfo{year}{1992}); \bibinfo{journal}{\prb}
  \textbf{\bibinfo{volume}{48}}, \bibinfo{pages}{10345}
  (\bibinfo{year}{1993}).}

\bibitem[{\citenamefont{{Schollw{\"o}ck}}(2005)}]{2005RvMP...77..259S}
\bibinfo{author}{\bibfnamefont{U.}~\bibnamefont{{Schollw{\"o}ck}}},
  \bibinfo{journal}{Rev. Mod. Phys.} \textbf{\bibinfo{volume}{77}},
  \bibinfo{pages}{259} (\bibinfo{year}{2005}).

\bibitem[{\citenamefont{Feiguin and Heidrich-Meisner}(2007)}]{feiguin:220508}
\bibinfo{author}{\bibfnamefont{A.~E.} \bibnamefont{Feiguin}} \bibnamefont{and}
  \bibinfo{author}{\bibfnamefont{F.}~\bibnamefont{Heidrich-Meisner}},
  \bibinfo{journal}{\prb} \textbf{\bibinfo{volume}{76}},
  \bibinfo{eid}{220508(R)} (\bibinfo{year}{2007}).

\bibitem[{\citenamefont{{Batrouni} et~al.}()\citenamefont{{Batrouni},
  {Huntley}, {Rousseau}, and {Scalettar}}}]{2007arXiv0710.1353B}
\bibinfo{author}{\bibfnamefont{G.~G.} \bibnamefont{{Batrouni}}},
  \bibinfo{author}{\bibfnamefont{M.~H.} \bibnamefont{{Huntley}}},
  \bibinfo{author}{\bibfnamefont{V.~G.} \bibnamefont{{Rousseau}}},
  \bibnamefont{and} \bibinfo{author}{\bibfnamefont{R.~T.}
  \bibnamefont{{Scalettar}}}, \bibinfo{howpublished}{arXiv:0710.1353v1}.

\bibitem[{\citenamefont{{Rizzi} et~al.}()\citenamefont{{Rizzi}, {Polini},
  {Cazalilla}, {Bakhtiari}, {Tosi}, and {Fazio}}}]{2007arXiv0712.3364R}
\bibinfo{author}{\bibfnamefont{M.}~\bibnamefont{{Rizzi}}},
  \bibinfo{author}{\bibfnamefont{M.}~\bibnamefont{{Polini}}},
  \bibinfo{author}{\bibfnamefont{M.~A.} \bibnamefont{{Cazalilla}}},
  \bibinfo{author}{\bibfnamefont{M.~R.} \bibnamefont{{Bakhtiari}}},
  \bibinfo{author}{\bibfnamefont{M.~P.} \bibnamefont{{Tosi}}},
  \bibnamefont{and} \bibinfo{author}{\bibfnamefont{R.}~\bibnamefont{{Fazio}}},
  \bibinfo{howpublished}{arXiv:0712.3364v1}.

\bibitem[{\citenamefont{{Olshanii}}(1998)}]{1998PhRvL..81..938O}
\bibinfo{author}{\bibfnamefont{M.}~\bibnamefont{{Olshanii}}},
  \bibinfo{journal}{\prl} \textbf{\bibinfo{volume}{81}}, \bibinfo{pages}{938}
  (\bibinfo{year}{1998}).

\bibitem[{\citenamefont{{Moritz} et~al.}(2005)\citenamefont{{Moritz},
  {St{\"o}ferle}, {G{\"u}nter}, {K{\"o}hl}, and
  {Esslinger}}}]{2005PhRvL..94u0401M}
\bibinfo{author}{\bibfnamefont{H.}~\bibnamefont{{Moritz}}},
  \bibinfo{author}{\bibfnamefont{T.}~\bibnamefont{{St{\"o}ferle}}},
  \bibinfo{author}{\bibfnamefont{K.}~\bibnamefont{{G{\"u}nter}}},
  \bibinfo{author}{\bibfnamefont{M.}~\bibnamefont{{K{\"o}hl}}},
  \bibnamefont{and}
  \bibinfo{author}{\bibfnamefont{T.}~\bibnamefont{{Esslinger}}},
  \bibinfo{journal}{\prl} \textbf{\bibinfo{volume}{94}},
  \bibinfo{pages}{210401} (\bibinfo{year}{2005}).

\bibitem[{LOp()}]{LOphasediagram}
\bibinfo{note}{A large error bar at $P=0.2$ takes into account the possibility
  of a reentrant region in which an LO condensate appears for intermediate
  values of $(k_\textrm{F}a_\textrm{1D})^{-1}$.}

\bibitem[{\citenamefont{{Machida} and {Nakanishi}}(1984)}]{1984PhRvB..30..122M}
\bibinfo{author}{\bibfnamefont{K.}~\bibnamefont{{Machida}}} \bibnamefont{and}
  \bibinfo{author}{\bibfnamefont{H.}~\bibnamefont{{Nakanishi}}},
  \bibinfo{journal}{\prb} \textbf{\bibinfo{volume}{30}}, \bibinfo{pages}{122}
  (\bibinfo{year}{1984}).

\bibitem[{\citenamefont{{Partridge}
  et~al.}(2006{\natexlab{b}})\citenamefont{{Partridge}, {Li}, {Liao}, {Hulet},
  {Haque}, and {Stoof}}}]{2006PhRvL..97s0407P}
\bibinfo{author}{\bibfnamefont{G.~B.} \bibnamefont{{Partridge}}},
  \bibinfo{author}{\bibfnamefont{W.}~\bibnamefont{{Li}}},
  \bibinfo{author}{\bibfnamefont{Y.~A.} \bibnamefont{{Liao}}},
  \bibinfo{author}{\bibfnamefont{R.~G.} \bibnamefont{{Hulet}}},
  \bibinfo{author}{\bibfnamefont{M.}~\bibnamefont{{Haque}}}, \bibnamefont{and}
  \bibinfo{author}{\bibfnamefont{H.~T.~C.} \bibnamefont{{Stoof}}},
  \bibinfo{journal}{\prl} \textbf{\bibinfo{volume}{97}},
  \bibinfo{pages}{190407} (\bibinfo{year}{2006}{\natexlab{b}}).

\bibitem[{\citenamefont{{de Silva} and {Mueller}}(2006)}]{2006PhRvA..73e1602D}
\bibinfo{author}{\bibfnamefont{T.~N.} \bibnamefont{{de Silva}}}
  \bibnamefont{and} \bibinfo{author}{\bibfnamefont{E.~J.}
  \bibnamefont{{Mueller}}}, \bibinfo{journal}{\pra}
  \textbf{\bibinfo{volume}{73}}, \bibinfo{pages}{051602}
  (\bibinfo{year}{2006}).

\bibitem[{\citenamefont{{Haque} and {Stoof}}(2007)}]{2007PhRvL..98z0406H}
\bibinfo{author}{\bibfnamefont{M.}~\bibnamefont{{Haque}}} \bibnamefont{and}
  \bibinfo{author}{\bibfnamefont{H.~T.~C.} \bibnamefont{{Stoof}}},
  \bibinfo{journal}{\prl} \textbf{\bibinfo{volume}{98}},
  \bibinfo{pages}{260406} (\bibinfo{year}{2007}).

\bibitem[{\citenamefont{Yang}(1962)}]{RevModPhys.34.694}
\bibinfo{author}{\bibfnamefont{C.~N.} \bibnamefont{Yang}},
  \bibinfo{journal}{Rev. Mod. Phys.} \textbf{\bibinfo{volume}{34}},
  \bibinfo{pages}{694} (\bibinfo{year}{1962}).

\end{thebibliography}
\end{document}